\documentclass[conference]{IEEEtran}
\IEEEoverridecommandlockouts

\usepackage{multirow}
\usepackage{balance}
\usepackage{graphicx}
\usepackage{tabularx}  
\newcolumntype{R}{>{\raggedleft\arraybackslash}X}
\newcolumntype{L}{>{\raggedright\arraybackslash}X}
\newcommand{\pln}[1]{\texttt{#1}}
\newcommand{\lab}[1]{\blue{\texttt{#1*}}}
\newcommand{\ftemp}{temp.py}

\usepackage{booktabs}
\usepackage{adjustbox}
\usepackage{framed}
\usepackage[table]{xcolor}
\usepackage{makecell}
\usepackage{url}
\usepackage{multicol}  
\usepackage{colortbl}
\usepackage{amsmath}
\usepackage[framemethod=TikZ]{mdframed}  
\usepackage{enumitem}
\usepackage{soul}
\usepackage{indentfirst} 
\usepackage{hyperref}
\usepackage{cleveref}
\usepackage{tcolorbox} 
\usepackage{xcolor}
\usepackage{subfigure}
\usepackage{calc}

\providecommand{\Description}[2][]{}

\newlength{\filecol}\newlength{\typecol}\newlength{\ncol}\newlength{\catcol}

\usepackage{framed}
\usepackage{xcolor}
\usepackage{fancybox}

\definecolor{lavback}{HTML}{E1D5E7}
\definecolor{darkpurple}{HTML}{4B2E5A}
\colorlet{lavfill}{lavback!70}

\newenvironment{takeawaybox}{%
  \def\FrameCommand{\fboxsep=8pt\colorbox{lavfill}}%
  \MakeFramed{\advance\hsize-\width \FrameRestore}%
}{%
  \endMakeFramed
}

\newcommand{\takeaway}[1]{%
  \begin{takeawaybox}%
  \small\noindent\textbf{\textcolor{darkpurple}{Takeaway:}} #1%
  \end{takeawaybox}%
}

\definecolor{interviewteal}{HTML}{E69F00}

\newenvironment{interviewquotebox}{%
  \def\FrameCommand{%
    {\color{interviewteal}\vrule width 2pt}%
    \hspace{5pt}%
    \colorbox{interviewteal!8}%
  }%
  \MakeFramed{\advance\hsize-\width \FrameRestore}%
}{%
  \endMakeFramed
}

\newcommand{\interviewquote}[3]{%
  \begin{interviewquotebox}
  \noindent\itshape\small ``#3''
   \; \normalfont\small\textcolor{interviewteal}{(\textbf{#1}, #2 years of experience)}
  \end{interviewquotebox}%
}

\newcommand{\revised}[3]{%
  \begingroup
  \def\reviewerColor{}%
  \ifnum#2=1\def\reviewerColor{black}\fi
  \ifnum#2=2\def\reviewerColor{black}\fi
  \ifnum#2=3\def\reviewerColor{black}\fi
  \textcolor{\reviewerColor}{#3}%
  \endgroup
}

\definecolor{myblue}{RGB}{0,70,140}

\newcommand{\blue}[1]{\textcolor{myblue}{#1}}

\begin{document}

\title{Same Scrutiny, More Time: Eye Tracking Insights into Reviewing LLM-Labelled Code}

\author{
Ranim Khojah$^1$, Francisco Gomes de Oliveira Neto$^1$, Mazen Mohamad$^{1,2}$, Julian Frattini$^1$, Philipp Leitner$^1$\\
$^1$\textit{Chalmers University of Technology and University of Gothenburg}\\ $^2$\textit{RISE Research Institutes of Sweden}\\
Gothenburg, Sweden \\
khojah@chalmers.se, francisco.gomes@cse.gu.se, mazen.mohamad@ri.se, \{julfrat,philipp.leitner\}@chalmers.se}

\maketitle

\begin{abstract}
Modern software development increasingly involves the use of large language models (LLMs) to generate code.
Despite their rapid advancement, LLMs remain prone to errors and hallucinations, emphasizing the importance of careful code inspection.
However, in practice, developers' trust in LLM-generated code and their willingness to review it thoroughly may not align with these recommendations, which is a gap that remains largely unexplored.
In this study, we conduct a Wizard-of-Oz experiment to examine how software engineers behave when code is explicitly labelled as LLM-generated during a code review task. We collect both behavioral data and participant feedback through eye-tracking and exit interviews.
Combining Bayesian data analysis with qualitative analysis, we found that while the thoroughness of code review did not change for participants, they spent more time fixating on LLM-labelled code,
indicating that the label itself influences attention.
Practitioners also adapted their review strategy for LLM-labelled code by assessing the code based on specific criteria (e.g., logical correctness), or using the prompt to guide their review. These findings inform LLM-based tool design on labelling while incorporating the prompt as a software artifact.
Our study reveals a gap between reviewers' intentions and actual reviewing behaviour, highlighting the need for software companies to revisit their AI policies (particularly regarding LLM-assisted development) to better support developers in reviewing LLM-generated code.

\end{abstract}


\section{Introduction}

In the current era of AI-assisted software development, developers have been relying on large language models (LLMs) and LLM-chatbots for code generation and modification~\cite{khojah2024beyond}, sometimes even reaching a high level of usage in what practitioners call vibe coding.\footnote{\url{https://www.ibm.com/think/topics/vibe-coding}} 
However, such use cases come with changes in the role of software developers. Particularly, code review becomes a central task to ensure that LLM-generated code is correct, secure, and compliant with established practices~\cite{lyu2025composer}. This is also consistent with many LLM policies in the software industry, where the validation of LLM-generated code and extra attention towards hallucinated content is becoming essential~\cite{khojah2026policy}.


Code review is not only a technical activity but also a social and cognitive process shaped by developers' motivations, experiences, and team dynamics~\cite{heander2025support}. Prior studies show that developers engage in code review for a range of goals beyond defect detection, including knowledge transfer, team awareness, and shared ownership~\cite{bacchelli2013expectations,heander2026code}. These interpersonal factors often rely on implicit trust, e.g., in teams where reviewers approve changes with expectations that authors will address issues without further scrutiny~\cite{heander2026code}. 

This raises an important question: how do such judgment- and experience-driven behaviours translate when the author is not a human, but an AI system? Developers may apply the same level of care and scrutiny to LLM-generated code, or adjust their behaviour based on prior experiences and perceptions of AI. Understanding these patterns can help teams design more effective review practices. This includes reinforcing consistent review strategies or addressing systematic differences in how code is assessed.

We conduct a Wizard-of-Oz (WoZ) experiment with 32 participants to investigate how software engineers behave when reviewing code that is either labelled as LLM-generated or left unlabelled, thereby shaping how participants \textit{perceive} its origin in a controlled setting.
We quantitatively analyse the gaze data collected from an eye-tracking device to examine how \textit{long} and \textit{thoroughly} participants review LLM-labelled code. Additionally, exit interviews capture the \textit{reviewing strategies} developers adopt when they perceive code as LLM-generated. Our research questions are:\\

\noindent \textbf{RQ1. To what extent does the presence of labels identifying code as LLM-generated affect how developers visually process the code during review?}

We use Bayesian data analysis (BDA) to compare the fixation durations and saccade (scan) lengths of participants when reviewing code labelled and not labelled as LLM code. 
We found that even though the level of thoroughness (saccade lengths) is comparable, the participants spent more time fixating on code that is labelled as LLM-generated (up to 60\% for more complex code).\\

\noindent \textbf{RQ2. What reviewing approaches do developers adopt when reviewing code perceived as LLM-generated?}

We qualitatively analysed the 32 gaze paths together with the exit interviews to understand how participants approached reviewing code perceived as LLM-generated with respect to their trust and assumptions. 
Many participants (n = 20, 62.5\%) reported no change in their evaluation criteria between labelled and unlabelled code, while 12 participants applied different evaluation criteria, such as placing greater emphasis on logical correctness or code quality. Independent of this, 14 participants used the prompt during their review, treating it either as a requirement to validate against or as contextual documentation.\\

This work makes several contributions: \textit{(i)} unlike previous research, we go beyond the quality of LLM-generated code and isolate the impact of \textit{perceived} code provenance (LLM vs. human) through our WoZ experimental design. Moreover, \textit{(ii)} our replicable experimental protocol can be adopted by software organisations to shape their AI policies regarding using LLMs in development and evaluate how their developers perceive and respond to LLM-generated code in practice; (iii) our findings suggest that the broader context of LLM-generated code (e.g., the prompt) acts as a software artifact that can support the design of LLM-integrated IDE features. Finally, \textit{(iv)} the shared scripts and data from our BDA \cite{replicationpackage} provide a robust and nuanced approach that accounts for individual differences across participants and \textit{(v)} act as prior data for future experiments.









\section{Eye Tracking in Software Engineering}

Eye tracking is a method used to record and analyse where and how long a person visually processes an artifact. Eye tracking studies in software engineering are typically built around four key concepts, as outlined by Grabinger et al.~\cite{grabinger2025cookbook}. The \textbf{stimulus} is the visual artifact presented to the participant. In code-related tasks, this is typically source code displayed in an integrated development environment (IDE) or a code review tool. The stimulus is then segmented into \textbf{areas of interest (AOIs)}, which are regions of the stimulus that the researcher wants to analyse separately, such as specific functions or code blocks. In turn, gaze data is collected through two primary measures, each capturing a different aspect of reading behaviour.

\textbf{Fixations} occur when the eye remains stationary at a point for more than 200 ms, indicating that the participant is actively processing (not just looking at) the content at that location. In code review, the total fixation duration within an AOI reflects \textit{how long} a developer spends reviewing a particular piece of code.
    
\textbf{Saccades} are rapid eye movements that connect two fixations, reflecting how the eye scans across the stimulus. Shorter saccades indicate a more detailed inspection, commonly observed when readers aim to understand the meaning of the stimulus (e.g., what the code does)~\cite{biedert2012readmeaning}, or process difficult content~\cite{rayner2009difficult}. In contrast, longer saccades indicate broader scanning and potential skipping. Consequently, in the context of code review, saccade length is a suitable proxy for the thoroughness of a reviewer's inspection.

Fixations and saccades are complementary and not correlated~\cite{rayner2009difficult}. Combining them (e.g., through gaze paths) reveals patterns of eye movements (e.g., transitions and order) that provide insights into visual exploration strategies and cognitive processing~\cite{grabinger2025cookbook}.

\section{Related Work}

As LLMs become increasingly common in software development, many companies adopt them in their development processes \cite{khojah2026policy}.
However, LLMs struggle with code-related tasks, particularly in terms of correctness, code quality, and other context-specific limitations~\cite{hou2024llmchallenges}. To manage these limitations, organisations often establish policies that govern the use of LLM-generated code~\cite{khojah2026policy}. More specifically, these policies often highlight the need for code review and verification when generating code, but do not specify \textit{how} the reviews should be conducted. 

Code review is a well-established practice in software engineering and serves as an essential step to detect bugs and improve code quality \cite{bacchelli2013expectations}. 
Beller et al.~\cite{beller2014changes} present a taxonomy of changes resulting from modern code review and show that the majority of these changes are evolvability changes (e.g., structure and documentation), while the remaining 15\% are functional changes (e.g., larger defects and logic).
In addition, code comprehension and understandability have been found to be a central challenge in effective code review~\cite{bacchelli2013expectations}, where reviewers often require specific information, such as code context and correct understanding, to conduct a thorough review~\cite{pascarella2018information}.
However, as LLM-generated code becomes increasingly prevalent in software companies, it remains unclear 
whether developers approach the review of LLM-generated code differently from human-written code.

Investigating this requires methods that go beyond self-reports and outcome measures. Eye tracking studies, in particular, have been widely used to investigate visual attention in software engineering tasks, especially for code comprehension~\cite{park2024eye}, program repair~\cite{tang2024validating}, and code review~\cite{chandrika2017traits, bertram2020trustworthiness}.
Chandrika et al.~\cite{chandrika2017traits} found that there is a correlation between the fixation duration on buggy code and the number of bugs found. Similarly, Park et al.~\cite{park2024eye} conducted an eye tracking study on code comprehension and found that logical correctness does not impact the total review time, but following a readability rule (minimise nesting) decreases the duration of fixations when reviewing the code. 

Prior work has examined how the perceived origin of code influences review behaviour. Bertram et al.~\cite{bertram2020trustworthiness} showed that labelling code as machine-generated alters both attention and trust, with reviewers shifting their gaze from code to tests when patches are presented as automated. Similarly, Tang et al.~\cite{tang2024validating} studied how developers validate and repair code generated by GitHub Copilot, finding that awareness of LLM-generated code leads to increased search effort and more frequent gaze transitions between code and comments. These studies suggest that code provenance plays a role in shaping how developers allocate attention during review.

Our work builds on these findings by focusing specifically on LLM-generated code in a controlled code review setting. Unlike prior work, which compares actual LLM-generated and human-written code, we employ a Wizard-of-Oz design in which all code is human-written but presented as either LLM-generated or not. This allows us to isolate the effect of perceived code provenance from differences in code quality, strengthening internal validity. Furthermore, beyond quantitative eye-tracking measures, we incorporate qualitative analysis of gaze paths and exit interviews to understand how reviewers interpret and act on these perceptions, revealing their reviewing strategies and underlying reasoning.
\section{Experimental Setup}

\begin{figure}[!ht]
    \centering
    \includegraphics[width=\linewidth]{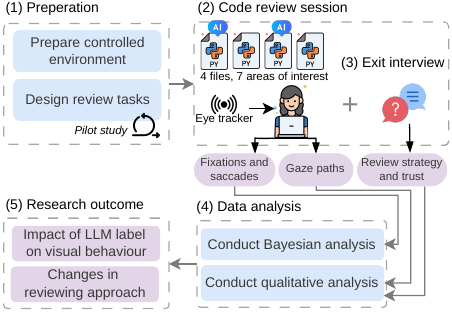}
    \Description[The process we followed in the study.]{The process we followed in this study.}
    \caption{The process we followed in the study}
    \label{fig:method}
\end{figure}

To understand how developers actually behave when reviewing LLM-generated code, we conducted a within-subject experiment with a crossover design, following the guidelines of Vegas et al.~\cite{vegas2016crossover}. The process we followed is illustrated in Figure \ref{fig:method}. We first set up the experimental environment by controlling unwanted factors that could disturb eye-tracking measurements, such as screen brightness and external lighting.
Then, we employed a WoZ approach~\cite{bernsen1994wizard} where 
we designed four pull requests (PRs) and applied LLM labels to selected code segments, where the entire code was, in fact, written by human developers (the wizard)~\cite{woz}. Label design and code segment selection were refined through a pilot study.

During a 1-hour session, we used an eye tracker while participants reviewed four Python files presented as PRs\footnote{No formal ethics committee approval was required to review the study as per the university’s guidelines and national regulations.}. Across the four PRs, two files contained one or two pieces of code labelled as LLM-generated, i.e., \textit{labelled areas of interest}. The order of pull requests and label assignments was rotated across participants so that the same areas of interest were reviewed both with and without a label by different participants. The session ended with an exit interview on the review approach and trust in LLM tools.

We applied Bayesian data analysis to the eye-tracking data (fixations and saccades) to assess whether the presence of an LLM label impacted the thoroughness and time participants spent reviewing the labelled code. We qualitatively analysed the interview data together with the gaze paths to understand \textit{how} participants adjusted their reviewing behaviour and which factors influenced these changes.

\subsection{Participants Selection}
We recruited 32 software engineering practitioners of different roles who work across 20 software organisations (see Table \ref{tab:participants_roles}). These organisations vary in size and domain and are located in three cities across two countries in Europe. When selecting the participants, we only invited practitioners who are familiar with Python and actively use version control and pull requests.


\begin{table}[ht]
\centering
\footnotesize
\caption{Number of participants for each role and average years of experience in the software industry.}
\label{tab:participants_roles}
\begin{tabular*}{\linewidth}{@{\extracolsep{\fill}}lrr@{}}
\toprule
\textbf{Role} & \textbf{N} & \textbf{Avg. years of exp.} \\
\midrule
Software Developer / Engineer & 20 & 7.0 \\
Test \& QA Roles & 4 & 3.6 \\
Data Scientist / Engineer & 2 & 6.0 \\
Research-Oriented Engineer & 2 & 13.5 \\
Game Developer & 2 & 3.3 \\
Security Engineer & 1 & 15.0 \\
DevOps Engineer & 1 & 9.0 \\
\midrule
\textbf{Total} & 32 & 7.0 $\pm$ 5.1 \\
\bottomrule
\end{tabular*}
\end{table}

\subsection{Defining Code Review Tasks}
Participants reviewed four Python files (on average, 77 $\pm$ 10 lines of code), each presented as a pull request in a review session that lasted up to 10 minutes. 
For simplicity, each pull request introduced a single, self-contained file with newly added code. We did not provide any tests to focus on participants' code comprehension and manual verification rather than their ability to run or debug tests.

\paragraph{Pull request selection} To ensure the pull request files are representative of real-world code, we base the files in the pull requests on Python code from two popular open-source projects, namely Requests\footnote{\url{https://github.com/psf/requests}}
(651 contributors, $\sim$10k forks; \texttt{storage.py} and \texttt{temperature.py}) and
Home Assistant\footnote{\url{https://github.com/home-assistant/core}} (4,646 contributors,
$\sim$37k forks; \texttt{sessions.py} and \texttt{utils.py}). To ensure that the code we use is entirely written by human developers, we used the official version of the repositories before the release of ChatGPT and GitHub Copilot (i.e., before October 2021). To ensure that review tasks can be completed within 10-minute time limit, we selected a subset of lines of code from each file to compose a new standalone file for the review task (e.g., we removed lines that included logging or invoked dependencies from other files). By avoiding dependencies across multiple files or mixed changes, we reduced potential confounding factors that might lead participants to prioritize certain parts of the code for reasons unrelated to the presence of the LLM label. 



\paragraph{Labelling code} 

The experiment includes \textit{one treatment} condition in which selected code segments, that is, AOIs, are labelled as LLM-generated, and \textit{one control} condition in which the same AOIs appear without the label. 
Depending on the participant, an AOI was either labelled as LLM-generated or left unlabelled. Across the four files, we defined seven AOIs of varying sizes (see Table~\ref{tab:aois}).


\begin{table}[!t]
\centering
\scriptsize
\setlength{\tabcolsep}{3pt}
\caption{Areas of interest, their corresponding file, and size in lines of code (LoC).}
\label{tab:aois}
\begin{tabularx}{\columnwidth}{l rrrr}
\toprule
 & \textbf{AOI 1} & \textbf{AOI 2} & \textbf{AOI 3} & \textbf{AOI 4} \\
\midrule
\textbf{File} & \texttt{sessions.py} & \texttt{sessions.py} & \texttt{storage.py} & \texttt{temper\-ature.py} \\
\textbf{LoC}  & 6 & 10 & 25 & 5 \\
\midrule
& & \textbf{AOI 5} & \textbf{AOI 6} & \textbf{AOI 7}  \\
\midrule
\textbf{File} & & \texttt{temper\-ature.py} & \texttt{utils.py} & \texttt{utils.py} \\
\textbf{LoC} &  & 4 & 2 & 21  \\
\bottomrule
\end{tabularx}
\end{table}

\begin{figure}[!ht]
    \centering
\includegraphics[width=\linewidth]{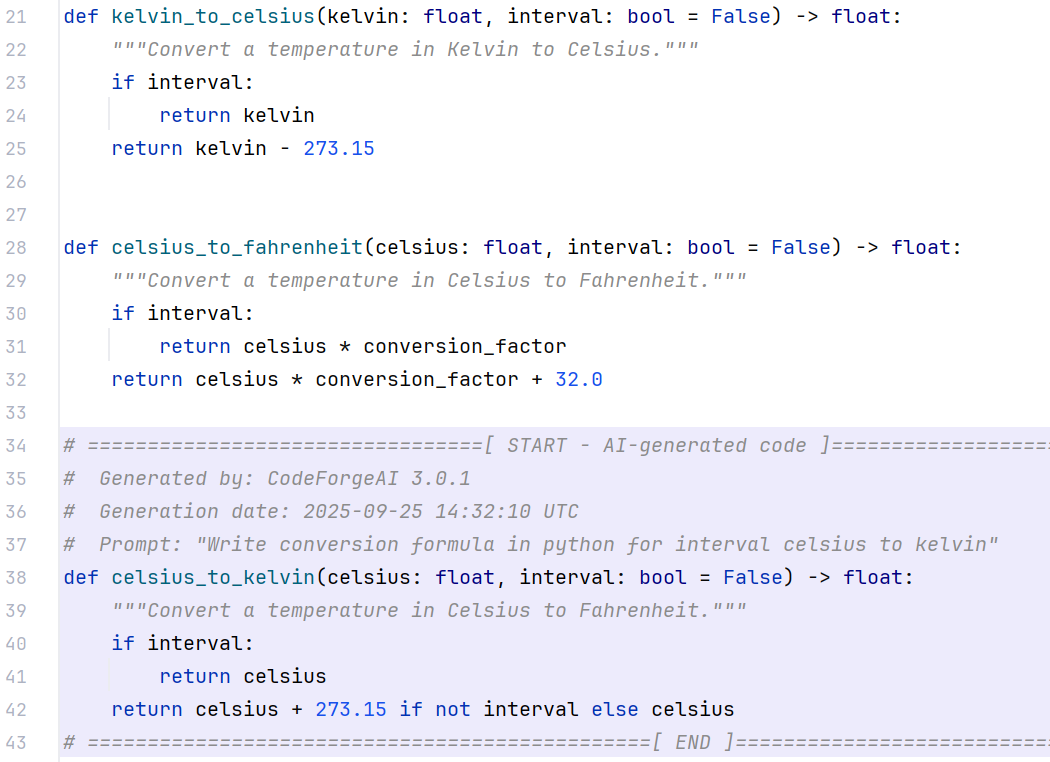}
        \caption{Standalone code in \texttt{temperature.py} (AOI 4, labelled)}
        \Description[Standalone code as a labelled AOI in \texttt{temperature.py}]{Standalone code as a labelled AOI in \texttt{temperature.py}}
    \label{fig:label-example}
\end{figure}

\begin{figure}[!ht]
    \centering
       \includegraphics[width=\linewidth]{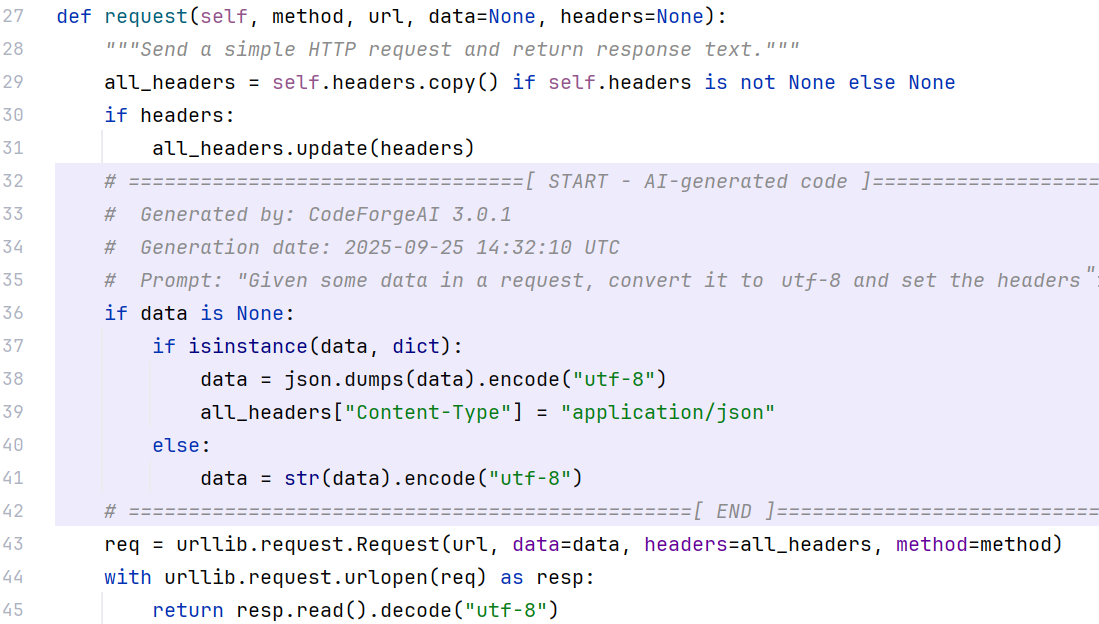}
        \caption{Inline code in \texttt{sessions.py} (AOI 1, labelled)}
        \Description[Inline code as a labelled AOI in \texttt{sessions.py}]{Inline code as a labelled AOI in \texttt{sessions.py}}
    \label{fig:label-example2}
\end{figure}

The selection of which segments receive the treatment (i.e., AOIs) was driven by covering different granularities, such as self-contained functions (Figure \ref{fig:label-example}) and individual lines within functions (Figure \ref{fig:label-example2}). Then, refinements of the selection of segments and the label format were informed by a pilot study with three Ph.D. students.
Their feedback helped identify code patterns that participants would plausibly perceive as LLM-generated.

The label consisted of a comment header indicating that the code was generated by a fictional LLM tool called CodeForgeAI, which we introduced to avoid biases stemming from participants' prior experiences with existing tools and to focus only on their perception of LLMs in general. To enhance realism, the label also included a generation timestamp and the prompt used to produce the code. The header and the associated code segment were highlighted in the IDE to visually distinguish them from the rest of the file.

\paragraph{Fault and code smell injection} To simulate realistic code review scenarios, we injected faults and code smells so that reviewers would encounter issues that could plausibly require attention during the review. 
We did not evaluate detecting these issues and informed participants that we would not assess their performance as it is outside the scope of the study.
The injected faults follow the categorizations of Python fault injection proposed by Marques et al.~\cite{marques2021injecting}, focusing on faults
that are syntactically valid but semantically incorrect code, such as using an incorrect variable, expression, or function call. For code smells, we followed established categories from a widely used catalogue.\footnote{\url{https://refactoring.guru/refactoring/smells}} Table~\ref{tab:faults-overview} shows the injected faults and code smell categories. These were distributed across labelled and unlabelled code segments to mitigate biases caused by unequal concentrations of issues.

\begin{table}[!ht]
\centering
\footnotesize
\caption{Overview of injected faults and code smells per file.}
\label{tab:faults-overview}
\settowidth{\filecol}{\texttt{temperature.py}}%
\settowidth{\typecol}{Code Smells}%
\settowidth{\ncol}{\textbf{N}}%
\setlength{\catcol}{\columnwidth-\filecol-\typecol-\ncol-6\tabcolsep}%
\begin{tabular}{@{}l l p{\catcol} c@{}}
\toprule
\textbf{File} & \textbf{Type} & \textbf{Categories} & \textbf{N} \\
\midrule
\multirow{2}{*}{\texttt{storage.py}}
  & Faults      & Checking, O-O Messages, Algorithm       & 3 \\
  & Code Smells & Speculative generality, Duplicated code & 2 \\
\midrule
\multirow{2}{*}{\texttt{temperature.py}}
  & Faults      & Initialization, Algorithm               & 3 \\
  & Code Smells & Dead code, Magic numbers                & 2 \\
\midrule
\multirow{2}{*}{\texttt{sessions.py}}
  & Faults      & Algorithm, Checking                     & 4 \\
  & Code Smells & Unneeded comments, Bad func. name       & 2 \\
\midrule
\multirow{2}{*}{\texttt{utils.py}}
  & Faults      & Checking, O-O Messages, Initialization  & 4 \\
  & Code Smells & Security risk, Bad function name        & 2 \\
\bottomrule
\end{tabular}
\end{table}


\paragraph{Order of pull requests} In our experimental design, we have 4 periods (individual review sessions) and 2 sequences (order in which treatment or control is presented), which resulted in 8 possible orders of the pull requests that we present in Table \ref{tab:groups}. The pull requests were arranged so that each pull request appears with both treatments in the same session: once with LLM-generated labels and once without, while ensuring that the two pull requests from the same project appear consecutively. For example, \texttt{storage.py} appears exactly twice in the first period; in \textbf{O1} with labelled code, and in \textbf{O5} without any labels. With 32 participants, this results in four participants being assigned to each sequence.

\begin{table}[!ht]
\centering
\scriptsize
\setlength{\tabcolsep}{4pt}
\caption{Pull request orders per participant. An asterisk (*) indicates labelled code. \texttt{storage.py} and \texttt{temperature.py} (abbreviated as \texttt{temp.py}) in Home Assistant; \texttt{sessions.py} and \texttt{utils.py} in Requests.}
\label{tab:groups}
\begin{tabularx}{\columnwidth}{@{}l llll@{}}
\toprule
\textbf{Ord.} & \textbf{Period 1} & \textbf{Period 2} & \textbf{Period 3} & \textbf{Period 4} \\
\midrule
O1 & \lab{storage.py}  & \pln{\ftemp}      & \lab{sessions.py} & \pln{utils.py} \\
O2 & \lab{\ftemp}      & \pln{storage.py}  & \lab{utils.py}    & \pln{sessions.py} \\
O3 & \lab{sessions.py} & \pln{utils.py}    & \lab{storage.py}  & \pln{\ftemp} \\
O4 & \lab{utils.py}    & \pln{sessions.py} & \lab{\ftemp}      & \pln{storage.py} \\
\midrule
O5 & \pln{storage.py}  & \lab{\ftemp}      & \pln{sessions.py} & \lab{utils.py} \\
O6 & \pln{\ftemp}      & \lab{storage.py}  & \pln{utils.py}    & \lab{sessions.py} \\
O7 & \pln{sessions.py} & \lab{utils.py}    & \pln{storage.py}  & \lab{\ftemp} \\
O8 & \pln{utils.py}    & \lab{sessions.py} & \pln{\ftemp}      & \lab{storage.py} \\
\bottomrule
\end{tabularx}
\end{table}

\subsection{Data Collection}
Each participant had to complete three main steps: pre-study questionnaire, code review session, and the exit interview. 

The \textbf{pre-study questionnaire} obtained written consent to participate in the study, with data anonymized, and allowed participants to opt out at any time. It collected information about the role and years of experience. It also included three Likert-scale items confirming participants' familiarity with Python and their confidence in writing and understanding Python code. The Likert-scale items were phrased as statements to which participants indicated their level of agreement.

The \textbf{code review session} took place in a controlled environment with fixed ambient lighting and screen brightness, using a two-screen setup. The main screen (2560 × 1140) displayed pull request files in PyCharm\footnote{\url{https://www.jetbrains.com/pycharm/}}, while participants wrote comments and selected their review decision according to GitHub pull request options (approve, comment, or request changes)\footnote{\url{https://docs.github.com/en/pull-requests/collaborating-with-pull-requests/reviewing-changes-in-pull-requests/about-pull-request-reviews}} on a secondary screen (1080 × 1920). We recorded eye movements using a Tobii Pro Spark eye tracker attached to the lower edge of the main screen at a frequency of 60 Hz, with gaze data captured via the CodeGRITS plugin~\cite{codegrits} and mapped to specific code elements.

The session began with eye tracker calibration, after which participants were briefed on their role and the task context: they were working in a company with an AI policy permitting the use of an internal chatbot (CodeForgeAI) for code generation, on the condition that the generated code is explicitly labelled in the codebase.

After reviewing all four pull requests, we conducted a 10-minute semi-structured \textbf{exit interview} with each participant covering their review strategies, their company's AI policy, and their trust and opinions on the use of LLMs for code-related tasks. Interview questions related to trust were based on Mayer et al.'s~\cite{mayer1995integrative} definition, which operationalises trust as the willingness to be vulnerable to another party's actions while perceiving risks posed by the trustee (in this case, LLM-generated code)~\cite{baltes2026rethinktrust}.


\subsection{Data Analysis}
\label{sec:analysis}

\paragraph{Bayesian data analysis (BDA)}


To infer the direction and strength of effects between relevant variables, we conduct BDA~\cite{mcelreath2018statistical} within a framework for statistical causal inference~\cite{siebert2023applications} based on Pearl's model of causality~\cite{pearl2009causality}. We choose BDA over frequentist methods (e.g., null hypothesis significance testing) because it provides richer insights~\cite{furia2019bayesian}, preserves uncertainty~\cite{mcelreath2018statistical}, and models underlying properties of the data~\cite{gren2021possible}, which is useful for analysing non-normally distributed data like fixation durations and saccade lengths in this study. Moreover, BDA coupled with our causal inference framework allows us to differentiate correlations from causal effects~\cite{pearl2009causality} and draw more reliable conclusions~\cite{furia2023towards}.

Our analysis follows a three-step process~\cite{siebert2023applications}: In the first \emph{modeling} step, we define our hypotheses and construct a directed, acyclic graph (DAG) representing variables (nodes) connected by (assumed) causal relationships. In our experiment, we formulate two hypotheses investigating the effect of the presence of the \emph{LLM label} on the total \emph{duration of fixations} and on the \textit{saccade lengths} within the areas of interest.

\begin{figure}[!ht]
    \centering
    \includegraphics[width=1\linewidth]{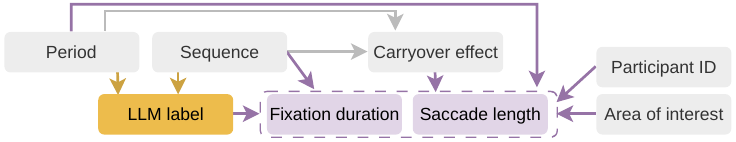}
    \caption{DAG for the impact of LLM label (yellow) on fixation duration and saccade length (purple), with covariates (grey).}
    \Description[DAG]{DAG for hypotheses regarding the impact of LLM label (yellow) on both the fixation duration and saccade lengths (purple), while accounting for other factors (grey).}
    \label{fig:dag}
\end{figure}

Our DAG (\Cref{fig:dag}) represents the relationships between the main variables of interest (``LLM label'', ``Fixation duration'', and ``Saccade length''), along with variables arising from the crossover design (i.e., \emph{period}, \emph{sequence}, and \emph{carryover}). It further includes covariates that may influence the outcome, such as the participant and the area of interest.

In the next \textit{identification} step, we determine which variables should be conditioned on to estimate the causal effect of interest while mitigating confounding~\cite{siebert2023applications}. In causal inference, conditioning refers to including variables in a regression model to block alternative paths through which bias could arise. Confounding occurs when a variable influences both the treatment and the outcome, potentially creating a spurious association between them~\cite{mcelreath2018statistical}. In our study, for example, the period (at what time a participant reviewed a particular AOI) can potentially affect participants' behavior in what is called the \emph{maturation} or \emph{exhaustion} effect~\cite{vegas2016crossover}, making it necessary to condition on it to avoid bias. Additionally, covariates such as \textit{area of interest} do not act as confounders but AOIs should still be included to improve the precision of the effect estimation by accounting for systematic variation across code segments~\cite{cinelli2024crash}. 

In the \emph{estimation} step, we formulate one regression model with prior distributions per hypothesis. Both outcome variables are positive quantities. We model the distribution of the outcome variable \emph{fixation duration} as a gamma distribution, as it emerges from the sum of individual measurements~\cite{mukherjee2010bayesian}. As the data-generating process of saccade lengths is less clear ontologically, we compare models using gamma and lognormal distributions and select the latter for its better fit~\cite{vehtari2017practical}.
Moreover, we include hierarchical effects~\cite{ernst2018bayesian} and interaction effects~\cite{aguinis2010best}. Hierarchical effects account for the nested structure of the data, such as observations grouped within users (\emph{user ID}). Interaction effects capture situations where the effect of one variable depends on another, for example, between \emph{LLM-label} and \emph{area of interest}, since the label influence may vary across different code segments.
The following formula shows the main parts of the Bayesian regression model for fixation durations.



{\footnotesize
\begin{align}
    \text{fixation} \sim{} & Gamma(\alpha, \beta) \label{eq:distribution} \\
    \log(\alpha) ={} & \delta + \delta_{\text{PID}}
        + \gamma_{Label} \, \text{Label}
        + \gamma_{AOI} \, \text{AOI} \label{eq:lm1} \\
    & + \tau_{\text{Label} \times \text{AOI}} \, \text{Label} \, \text{AOI} \label{eq:lm11} \\
    & + \gamma_{Seq} \, \text{Sequence}
        + \gamma_{P} \, \text{Period}
        + \gamma_{CO} \, \text{Carryover} \label{eq:lm2} \\
    \beta ={} & Gamma(10, 0.5) \label{eq:prior}
\end{align}
}

Equation~\ref{eq:distribution} models the total fixation duration using a Gamma distribution parameterized by a shape parameter $\alpha$ and a rate parameter $\beta$, which is suitable for modeling positive, right-skewed data such as fixation times. 
In summary, the model estimates how the presence of the LLM label and other factors influence fixation duration, while accounting for differences between participants and the structure of the experiment. 

The shape parameter $\alpha$ (Equations~\ref{eq:lm1} and \ref{eq:lm11}) is modeled as a function of our factors (on a log scale to ensure positive values). 
The model includes a global intercept ($\delta$) and participant-specific varying intercepts ($\delta_{\text{PID}}$) to account for individual differences in baseline fixation behaviour. 
We then include the main effects of the LLM label ($\gamma_{Label}$), the area of interest ($\gamma_{AOI}$), and their interaction ($\tau_{\text{Label} \times \text{AOI}}$). 
Finally, we adjust for sequence, period, and carryover effects (Equation~\ref{eq:lm2}) to account for the crossover design~\cite{vegas2016crossover}. 
Positive coefficients ($\gamma_x$) correspond to longer fixation durations.

Equation~\ref{eq:prior} specifies a prior for the rate parameter $\beta$ using a Gamma distribution. 
Similar weakly informative priors are assigned to all other coefficients ($\delta$, $\gamma$, and $\tau$)~\cite{wesner2021choosing}, but are omitted for brevity. 
Afterwards, we perform prior predictive checks to confirm their appropriateness~\cite{mcelreath2018statistical}. Once assessing that the training process converged accordingly~\cite{gelman1992inference} and no diagnostic suggested issues~\cite{mcelreath2018statistical}, we evaluate the model. The details of our BDA workflow, along with scripts and explanations, are in our replication package.


\paragraph{Qualitative Analysis}

We complement the statistical results with a qualitative analysis of the exit interviews, alongside gaze path visualizations showing how participants moved across lines of code during each pull request review. 
Three authors jointly analysed three interviews in an initial session and established initial themes related to trust, reviewing strategy, and self-reported behavioural changes. The remaining interviews were analysed by four authors across six sessions. Each session involved two authors, where any disagreements were discussed and resolved on the spot. As the analysis progressed, more concrete codes emerged around how participants used the label and criteria they applied during review.

\section{LLM Label Impact on Reviewing Behaviour}
\label{sec:results-bda}

Here, we report the results of our Bayesian data analysis on how the presence of a label indicating LLM-generated code affects review duration and thoroughness, measured through duration of fixations and saccade lengths. The Bayesian model produces posterior estimates and 95\% credibility intervals (CIs) for the effects of our factors ($\gamma_{\text{Label}}$), namely LLM label, the area of interest, and their interaction $\tau$. Recall that our model also accounts for period, sequence, and carryover effects (see Section \ref{sec:analysis}). An effect is considered credible when its CI \textit{does not} include zero. Positive estimates indicate longer fixation duration or longer saccade length when receiving the treatment, i.e., seeing the ``LLM-generated''-label, whereas negative estimates indicate shorter values.

For fixation duration, the effect of the LLM label is credible ($\gamma_{Label}$: 95\% CI = [0.09, 0.56], $\mu = 0.33$) as the intervals of the posterior distributions do not overlap with zero. The estimate ($\mu$) is positive, indicating that participants spent \textit{more} time fixating on code when labelled as LLM-generated compared to identical unlabelled code. In contrast, the period and carryover effects are not credible. This suggests that file review order and prior exposure to labelled code did not influence fixation durations, supporting the validity of the counterbalanced design.

\begin{figure}[!ht]
    \centering
    \includegraphics[width=1\linewidth]{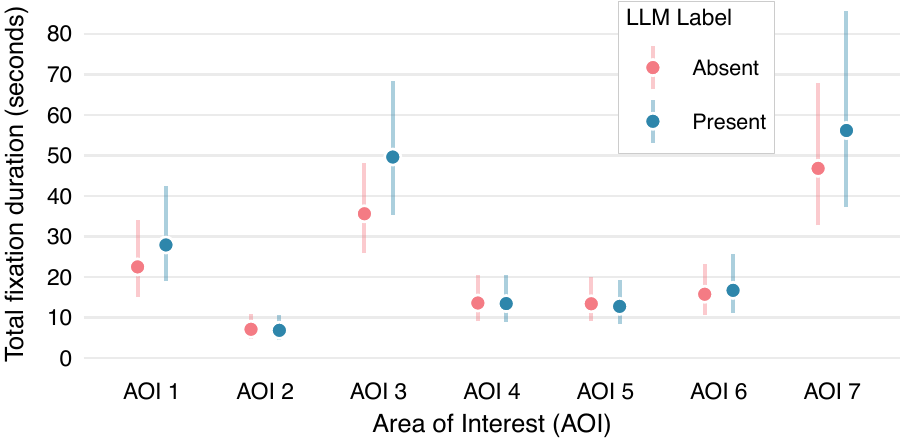}
    \caption{The effect of the presence of an LLM label on the total fixation duration per area of interest.}
    \label{fig:aoi-ai-fix}
\end{figure}

The interacting effect between the LLM label and the AOIs (Figure~\ref{fig:aoi-ai-fix}) shows where in the code the label effect is most pronounced. First, the CIs are large due to individual differences between participants (e.g., with regard to their reading speed), which is expected.

Second, review times for labelled code are either comparable to or longer than those for the same unlabelled code. This difference is more pronounced (up to 15 seconds) for code that requires more time to review (more than 20 seconds) and is negligible otherwise.

In practice, while differences in seconds may seem minor, they are substantial in relative terms, especially since they concern fixation time alone, not the full time spent on review. For instance, reviewing 6 lines of code in \texttt{sessions.py} (AOI~1) with an average time of 15 seconds, an additional 5 seconds is approximately a 33\% increase in reviewing duration. For more complex code, as in AOI~3 in \texttt{storage.py} (25 lines of code), reviewers spent an extra 15 seconds, a 60\% increase.
Our results suggest that there is no meaningful difference when reviewing very short code, but the time tax imposed by reviewing LLM contributions grows significantly the more complex the code becomes. Given that contemporary LLMs are able to generate pull requests of substantial size, including generating full files or making simultaneous non-trivial changes in multiple places in the code base~\cite{watanbe2026agentic, jimenez2023swe}, this finding is concerning.


Note that spending more time reviewing a piece of code in general is not necessarily due to the size of the code but rather its complexity. For example, AOI~4 in \texttt{temperature.py} (in Figure \ref{fig:label-example}) and AOI~1 in \texttt{sessions.py} (in Figure \ref{fig:label-example2}) are of comparable length (5 vs 6 lines of code). However, AOI~4 is simpler and follows the same pattern and style as in previous functions in the same file, while AOI~1 involves more complex and nested conditions.
This also aligns with what some participants mentioned in their interviews: if they encountered a complex LLM-generated code, their first intuition was that the complexity was unnecessary. 

\interviewquote{game developer}{2.5}{If there's some really odd code --- not necessarily bad, just strangely complex --- if a human wrote it, I'd think: okay, they went through the effort of writing this, so there's probably a reason behind it (...) But if it's an AI-written function, it immediately jumps out at me like: oh, this is strange}

Some participants also acknowledged that the margin of faults that LLMs produce grows with the size of the code. 

\takeaway{Practitioners spend more time fixating and reviewing code that is labelled as LLM-generated. This effect grows stronger in more complex code since it is expected that the complexity of LLM-generated code may be unnecessary.}

To complement the fixation duration results, we examine saccade lengths, which capture how developers scan the code. Shorter saccades indicate a more thorough ``line-by-line'' review, while longer saccades indicate broader scans. 
Most of our participants described in the exit interview that their general reviewing strategy consisted of going over the code in a top-down manner, then following the execution path when they encountered a called function.
Many also mentioned that they reviewed LLM-labelled code more \textit{thoroughly} even when they did not show a strong mistrust of LLMs.

\interviewquote{software engineer}{2.5}{It was good to mark that [the code] was AI-generated. And, of course, I approach it a bit more thoroughly, although, in my experience, [LLMs] are performing very well in real life}

\begin{figure}[!ht]
    \centering
    \includegraphics[width=1\linewidth]{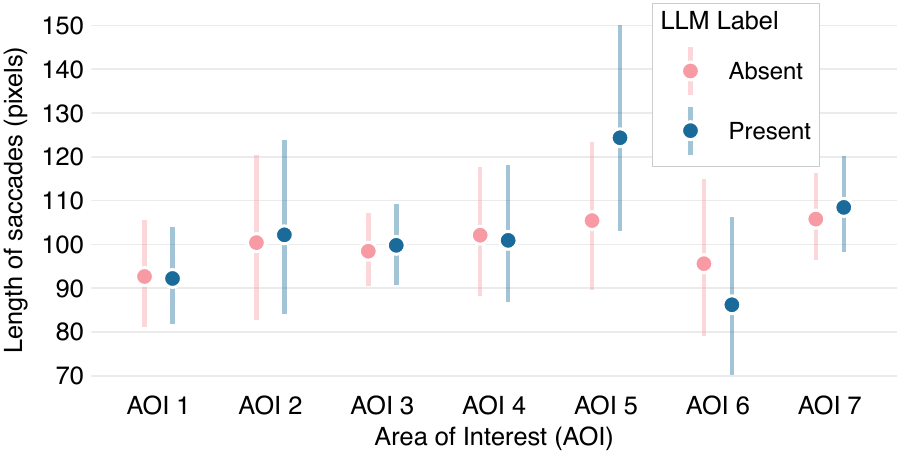}
    \caption{The effect of the presence of an LLM label on the saccade lengths (in pixels) per area of interest.}
    \Description[Saccades per AOI]{The effect of the presence of an LLM label on the saccade lengths (in pixels) per area of interest.}
    \label{fig:aoi-ai-sacc}
\end{figure}

However, the gaze data we collected revealed that the presence of an LLM-label does not change the length of saccades within an area of interest ($\gamma_{\text{Label}}$: 95\% CI = [-0.07, 0.10], $\mu = 0.01$). The CI overlaps with zero, indicating a non-significant effect of the LLM labels.
This is also apparent in Figure~\ref{fig:aoi-ai-sacc}, showing the impact of the label on saccade lengths per area of interest. 

While the effect is comparable for most areas of interest, with a difference of 2-3 pixels, we notice gaps in the opposite directions for the two shortest codes (AOI~5 in \texttt{temperature.py}, 4 lines, and AOI~6 in \texttt{utils.py}, 2 lines). Although there is no clear rationale behind the difference in saccade lengths, the 20–25 pixel difference we observe corresponds roughly to the length of a single word, which we argue is not significant in practice, and unlike the fixation duration, this effect does not appear to scale with code complexity or length. This also aligns with the work by Bertram et al.~\cite{bertram2020trustworthiness} on average saccade lengths when reviewing machine-generated code as opposed to human-written code. There were no significant differences in the saccades, and they were around 100 pixels on average.

Therefore, when our participants reported a \textit{more} careful or thorough review process, this reflected mainly in the total fixation duration (i.e., review time) rather than a more detailed review. The entire code (whether labelled or not) was reviewed in approximately the same thoroughness (with 85-110 pixel length scans). 


\takeaway{Practitioners review LLM-labelled code with the same thoroughness as unlabelled code. Participant statements regarding more careful reviews of LLM contributions appear to translate into longer reviewing time, but not more thorough reviews.}


\section{Changes in Reviewing Behaviour}
\label{sec:strategies}

The quantitative results show that developers spent more time fixating on code labelled as LLM-generated. 
To understand what drives this pattern and why the gap between self-reported careful review and actual gaze behaviour exists, we qualitatively analysed two data sources for all 32 participants: (i) gaze timelines from the review sessions and (ii) the exit interviews. By cross-referencing these sources, we examined how participants approached the review task, which beliefs and evaluation criteria shaped their judgments, and how the perceived provenance of the code, that is, whether it was believed to come from an LLM or a human developer, influenced their attention, reviewing strategy, or inspection criteria.

\subsection{Attention to LLM-Labelled Code}

Many participants (\textbf{n=20, 62.5\%}) reported that their \textbf{reviewing approach and criteria did not change} between labelled and unlabelled code. As a result, they described applying the same reviewing standards regardless of how the code was presented. 

\interviewquote{software engineer}{12}{Pretty soon, I just decided that it's all code. I need to review it all quite carefully anyway, so I treated it as any other code}

However, even participants who reported no change in strategy still tended to briefly pay attention to the prompt or LLM-labelled segments. This additional attention was not described as a deliberate strategy, but rather as a quick, instinctive check for obvious issues. In their interviews, these participants often expressed a general mistrust of LLMs, possibly reinforced by previous experiences with LLMs and the fact that their companies' AI policies are broad and offer little concrete guidance on the use of LLM-generated code. 

\interviewquote{software engineer}{8}{I would take a glance to see if it is doing something absolutely stupid, or else, I would depend on the test cases}

This suggests that, although participants perceived their reviewing approach as consistent, the presence of the LLM label subtly influenced their attention.

\takeaway{Participants following a consistent reviewing strategy still showed an unconscious attentional pull toward LLM-labelled code, highlighting a gap between stated reviewing intent and actual behaviour. This gap appears to be shaped by prior experiences and underlying mistrust of LLM-generated code.}

\subsection{Shifts in Evaluation Criteria}

Some participants (\textbf{n=12, 37.5\%}) adjusted the \textbf{evaluation criteria} applied to LLM-labelled code. Rather than reviewing all code using the same standards, these participants adjusted what they looked for depending on whether the code was perceived as LLM- or human-generated. These shifts in criteria were closely tied to participants' expectations about the types of mistakes made by LLMs versus human developers. In particular, trust and prior experience with AI played a central role in shaping these expectations.

Participants who expressed relatively higher trust in AI tended to expect fewer logical errors in LLM-generated code, but more issues related to code quality or consistency. As a result, their review focus shifted toward aspects such as readability, adherence to coding conventions, and consistency with the rest of the codebase (e.g., use of variables, libraries, or style), rather than strict logical correctness. Despite this shift in criteria, their overall reviewing flow often remained similar across labelled and unlabelled files, typically following a top-down inspection of the code.

\begin{figure}[!ht]
    \centering
    \includegraphics[width=1\linewidth]{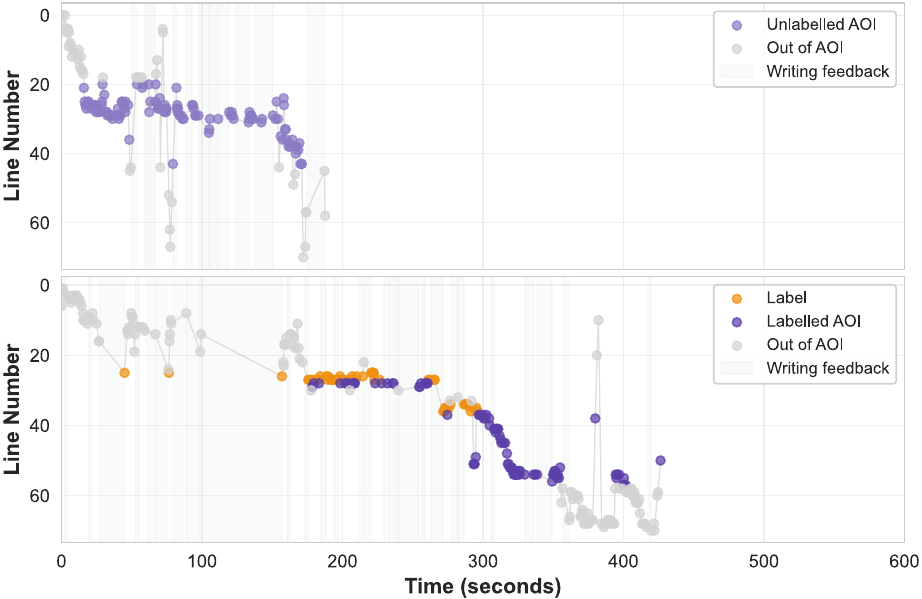}
    \caption{Gaze paths for a software test manager (6 years experience, O5) reviewing \texttt{storage.py} (unlabelled, top) and \texttt{utils.py} (LLM-labelled, bottom).}
    \Description[Gaze paths for a software test manager (6 years experience, O5)]{Gaze paths for a software test manager (6 years experience, O5) reviewing \texttt{storage.py} (unlabelled, top) and \texttt{utils.py} (LLM-labelled, bottom).}
    \label{fig:b118-stor-utils}
\end{figure}

For example, a test manager with 6 years of experience mentioned that when they see LLM-generated code, they trust it more than human-written code. This is because they expect to see more logical errors in human-written code, but still believe that \textit{``AI makes quality problems more than the humans''}. Therefore, they mentioned that they did not pay much attention to the LLM-labelled code unless they had extra time. While individual gaze paths varied, this participant's gaze (Figure~\ref{fig:b118-stor-utils}) reflects the general tendency among higher-trust participants: a similar reviewing flow across labelled and unlabelled files, moving from top to bottom through the code, but with more time spent on LLM-labelled segments.

In contrast, participants with lower trust in AI expected LLM-generated code to contain more logical errors. These expectations were often based on prior experiences with LLM-based tools, which they described as prone to misunderstanding prompts and producing semantically incorrect code. As a result, their review focused more strongly on verifying logical correctness and reasoning about the behaviour of the code.

\begin{figure}[!ht]
    \centering
    \includegraphics[width=1\linewidth]{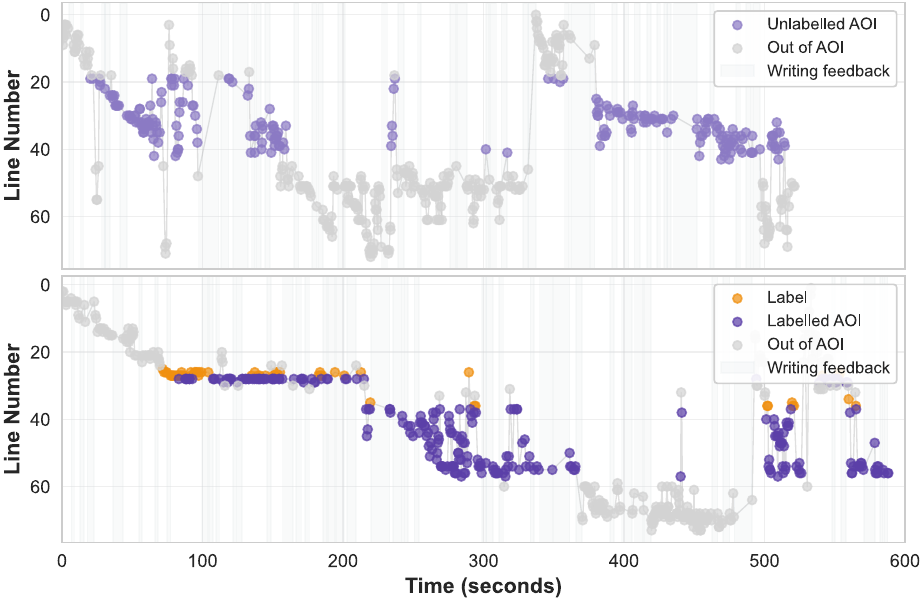}
    \caption{Gaze paths for a software developer (2 years experience, O5) reviewing \texttt{storage.py} (unlabelled, top) and \texttt{utils.py} (LLM-labelled, bottom).}
    \Description[Gaze paths for a software developer (2 years experience, O5)]{Gaze paths for a software developer (2 years experience, O5) reviewing \texttt{storage.py} (unlabelled, top) and \texttt{utils.py} (LLM-labelled, bottom).}
    \label{fig:b102-stor-util}
\end{figure}

Figure~\ref{fig:b102-stor-util} illustrates the gaze paths for a software developer who inspected the LLM-labelled code for logical errors. They believe that the logical errors that a human developer can make are usually captured by the tests. However, they believe that \textit{``the range of mistakes [LLMs] do is much broader than a human''}, and many of them can go undetected by unit tests and linters.

\interviewquote{software developer}{2}{The type of problems that a developer might do might get caught in other things like linting or test cases, but I believe that the AI-generated might pass those but still be logically wrong}

One reason that some participants described is that such faults are more likely to occur at the integration level rather than at the unit level. In the \texttt{utils.py} file (bottom), the developer spent around two minutes reviewing the first labelled code segment (gaze spanning between seconds 80--200). They also revisited the second labelled segment later in the review (around line 37 at second 500), whereas in the unlabelled \texttt{storage.py} file (top), they instead returned to the beginning of the file (line 1 at second 350).

Across these participants, we did not observe a consistent gaze pattern that clearly distinguishes those who changed their evaluation criteria from those who did not. However, the gaze data suggest that differences in expectations influence how attention is allocated within the file, for example through revisits or targeted inspection of specific segments. Our findings suggest that changes in reviewing behaviour are not uniform, but are shaped by participants' expectations of the strengths and limitations of LLM code.

\takeaway{Reviewers adjusted what they look for in LLM-labelled code. Depending on their trust in AI, they prioritize logical correctness or code quality, showing that prior expectations shape evaluation criteria, even when the overall review process remains similar.}

\subsection{Interpreting and Using the Prompt}

Now, we focus on how participants, regardless of the evaluation criteria, interact with the \textbf{LLM label itself}, specifically, the prompt embedded in the label. Rather than acting solely as a marker of code provenance, the prompt provides additional information that participants incorporate into their review. From the interviews, among those who reported using the prompt in their review (n=14), two groups emerged: (i) participants who treated the prompt as a \textbf{requirement description} (n=5), and (ii) those who treated it as \textbf{documentation} (n=7).
Two participants reported using both. 

These two interpretations reflect different ways of using the prompt during a review. When treated as a \textbf{requirement}, the prompt becomes an external reference against which the code is validated, shifting the reviewer's focus to \textit{alignment}. In this case, reviewers implicitly ask: \textit{``Did the LLM produce what was asked?''} In contrast, when treated as \textbf{documentation}, the prompt serves as contextual information to support understanding, shifting the focus to \textit{comprehension}, asking: \textit{``What is this code doing?''}.

Participants who treated the prompt as a requirement adjusted their review behaviour accordingly. Rather than reviewing the code in isolation, they repeatedly compared the code against the prompt to verify whether the generated output matched the intended specification. This resulted in a characteristic back-and-forth gaze pattern between the label and the corresponding code segments.

\begin{figure}[!ht]
    \centering
    \includegraphics[width=1\linewidth]{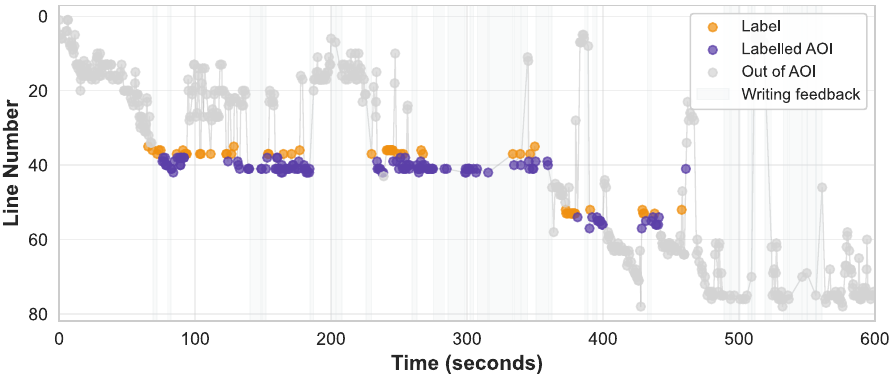}
    \caption{Gaze path in \texttt{temperature.py} (software engineer, 10 years experience, O7).}
    \Description[Gaze path in \texttt{temperature.py} (software engineer, 10 years experience, O7).]{Gaze path in \texttt{temperature.py} (software engineer, 10 years experience, O7).}
    \label{fig:b326-temp}
\end{figure}

Figure~\ref{fig:b326-temp} illustrates this behaviour for a software engineer with 10 years of experience. The participant frequently shifts between the prompt (yellow) and the labelled code (purple), indicating an ongoing validation process. 
This gaze pattern was consistent across participants in this group and reflects a shift in the nature of the task: from understanding the logic of the code to verifying whether the code generation \textit{worked} as intended. 
Which is in line with prior observations by Tang et al.~\cite{tang2024validating}, where developers similarly shifted between prompt and code to resolve mismatches between intended and generated behaviour.
In some cases, this also led our participants to 
question whether discrepancies were due to incorrect implementation or unclear intent of the author of the pull request. 

\interviewquote{software engineer}{1.5}{I was pointing out the discrepancy and telling them that: oh, did you actually intend to have this code, or is there some discrepancy between what you said and what AI generated for you?}

In contrast, participants who treated the prompt as documentation used it to support their understanding of the code, rather than a reference for validation. Their review behaviour typically followed a \textbf{label-first pattern}, where they first read the prompt to form an understanding of the functionality, then inspected the code.

\begin{figure}[!h]
    \centering
    \includegraphics[width=1\linewidth]{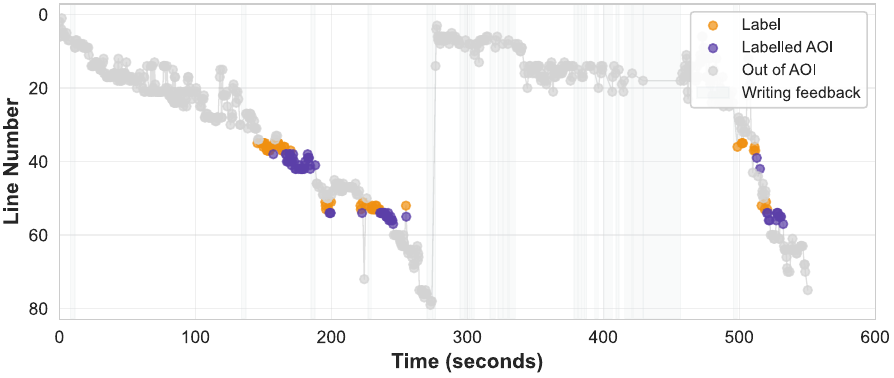}
    \caption{Gaze path in \texttt{temperature.py} (security engineer, 15 years experience, O2).}
    \Description[Gaze path in \texttt{temperature.py} (security engineer, 15 years experience, O2).]{Gaze path in \texttt{temperature.py} (security engineer, 15 years experience, O2).}
    \label{fig:a211-temp}
\end{figure}

Figure~\ref{fig:a211-temp} shows an example of this behaviour for a security engineer with 15 years of experience. Unlike the requirement-based pattern, this participant exhibits fewer shifts between label and code. Instead, they fixate on the prompt first (yellow), followed by a cluster of fixations on the labelled code (purple), suggesting a sequential process of understanding followed by inspection. In this case, the prompt was treated as a code artifact and was assessed with the rest of the code.

\interviewquote{software developer}{8}{I wrote that the prompt is unclear (...), usually AI makes mistakes when the prompt is unclear, and it makes things up}

While the number of passes through the file varied, this label-first strategy was common among participants in this group.

\takeaway{When reviewers use the prompt during code review, they adopt two approaches: as a requirement to validate alignment between intent and code, or as documentation to support code comprehension. These interpretations shift the focus of review from code alone to the relationship between code and its perceived origin.}


    
    
    
    
\section{Implications}

\paragraph{Trusting LLMs takes a different form than trusting human developers}
We learned that the relationship between trust and the reviewing behaviour is asymmetric. In other words, practitioners who trust LLMs do not necessarily spend less time reviewing LLM-generated code, nor do they review it less thoroughly. For example, one of the participants who expressed a strong trust towards LLMs is a software test manager (gaze paths in Figure~\ref{fig:b118-stor-utils}). Despite their trust, they still reviewed the two unlabelled PRs (presuming they have no LLM-generated content) in less time than the two with labelled codes. However, they focused on reviewing the labelled code in terms of quality over logical correctness.

On the other hand, most of the participants expressed mistrust towards LLMs at varying levels. This resulted in a difference in the behaviour (in terms of review time) \textit{only} for more complex code snippets (see Figure \ref{fig:aoi-ai-fix}) with no big difference in the saccade lengths, that is, the thoroughness of the review. In fact, many of them reported that they did not treat LLM-labelled code any differently and followed the same reviewing strategy they used for code written by human developers, just while being ``careful''. These differences in behaviour, despite mistrust, support the claims that acceptance of generated software artifacts does not reflect trust~\cite{baltes2026rethinktrust}.


However, measuring self-reported trust and carefulness is difficult, and research should not use them solely as a proxy to predict behaviour. Combining them with objective behavioural measures, e.g., eye tracking, can help capture the gap between how carefully developers think they are reviewing vs. how they actually are. Better operationalisation of trust in experiments is still needed; for instance, accounting for interpersonal familiarity, as some participants reported reviewing PRs from known or more senior colleagues less thoroughly, which is a factor to capture in future work.

\paragraph{AI Policies need to define the LLM-code verification process}
Reflecting on the different reviewing approaches that the participants followed for LLM-labelled code (see Section~\ref{sec:strategies}), we notice a gap between organisational policies and individual practices. While software organisations set AI policies that mandate a more ``careful'' verification of LLM-generated code, there is no specification of what this means in practice. This has resulted in software engineers merely improvising cautiousness. Sometimes, this meant checking if the code aligns with the intent of the developer, and other times it reflects in focusing on a specific aspect related to LLMs' limitations, e.g., code quality. However, for many, they do not review LLM-generated code any differently than any human-written code, even if they see the need to ``be extra careful.''

Similarly, the gaze data showed that while developers reviewed LLM-generated code as thoroughly as human-written ones, they planned their time for reviewing such code based on perceived complexity.
Therefore, AI policies need to define a structured verification process, including constraints on the \textit{size} of LLM-generated code (e.g., discouraging entire-file generation) and specifying \textit{when} it is submitted for review; ideally early in development, before blocks grow large and complex.
This is particularly important since research envisions a change in the role of a software developer to focus more on code verification~\cite{ferino2025walking, Simkute04032025}. Such needs were also highlighted by Nahar et al.~\cite{nahar2025comfort} regarding standardising responsible AI evaluation to mitigate ethical and legal issues for organisations. The process should also clarify responsibilities, as our participants were torn about whether the responsibility of verifying the LLM-generated code lies with the original developer or the reviewer.

\paragraph{The prompt is an artifact that should be available on-demand}
Although our experiment focused on marking code as LLM-generated, the label we designed also included additional metadata such as the LLM used, the date and time of generation, and the prompt (mainly to make the label feel realistic).
However, we found that these details, particularly the prompt and the LLM used, were unexpectedly important to many participants, revealing that prompt-to-code traceability is valuable during code review.

More specifically, our participants expressed the importance of knowing the original intent behind the generated code, especially because LLMs do not always produce code that clearly reflects what was asked. The prompt was also useful to understand LLM-generated code that tends to be unnecessarily complex. 
A similar behaviour was observed by Tang et al.~\cite{tang2024validating} when developers repaired code generated by GitHub Copilot. Consequently, we argue that there is a need to add mechanisms to code generation and review tools to map code to the prompt(s) used to generate it, ideally in a way that does not disrupt the reviewing flow.

To improve the usability of such a tool, we encourage LLM-based tool designers to keep the prompt-to-code traceability available within the development environment and on-demand to give reviewers access to the context they need, without adding noise for those who do not need it. 
We also learned that storing prompts as code comments or any code artifact is not ideal (this may, for example, conflict with company coding practices or clean code principles~\cite{martin2009clean}). Instead, prompts should be treated as a separate artifact that can be stored and accessed as code metadata. This is particularly relevant for multi-turn interactions, where the full conversation history (not just a single prompt) needs to be preserved.

\section{Validity Threats}

The crossover design induces threats to \textbf{internal validity}~\cite{vegas2016crossover}, e.g., the learning effect over multiple periods or an optimal sequence effect.
We mitigate these by adjusting for them in our analyses~\cite{vegas2016crossover,frattini2024crossover}, which not only factors out the effects from the phenomenon of interest but also allows us to determine that they did not affect our experiment.
In addition, to reduce the risk of hypothesis guessing, participants were invited to take part in a code review study without being informed that LLM-generated code was involved.

The \textbf{external validity} may be threatened by missing unobserved variables.
For example, the participants' stance towards LLM-generated code may affect their trust and, hence, moderate the effect of the LLM-label.
Consequently, the quantitative results may not generalize to a more or less trusting sample of subjects.
We are unable to mitigate the threat from these unmeasurable moderators. 
Therefore, we define our DAG (see Figure~\ref{fig:dag}) to scope our causal assumptions explicitly and confine it to observable variables.

Additionally, participants may behave differently under eye-tracking equipment than in natural settings.
This is an inherent trade-off of controlled experiments needed to isolate the effect of interest.
To reduce obtrusiveness, participants reviewed in a quiet room with one author seated facing away.
Moreover, while the PRs used in the study may not fully reflect real-world code review environments, we selected code from large and active open-source projects to ensure representation of collaborative development that we observe in the software industry. However, participants still reviewed a reduced version of these code files and were described as entirely new, whereas real PRs typically contain a mix of added, deleted, and modified lines. Despite this, such a scenario is not unrealistic, as some introduce entirely new modules or features.

To support \textbf{construct validity}, we use well-established eye-tracking metrics that capture visual attention \cite{grabinger2025cookbook}. We triangulate our analysis by combining quantitative measures, gaze paths, and qualitative data from exit interviews. 

Any subjective step in the BDA, e.g., DAGs, distribution family for outcome variables, and prior probability distributions, implies a potential threat to \textbf{conclusion validity}.
We mitigate them by following established principles, i.e., domain expertise pooling~\cite{glymour2008causal}, maximum entropy criterion~\cite{jaynes2003probability}, and uninformative priors~\cite{wesner2021choosing}.
\section{Conclusion }

Prior research has often argued that LLM-generated code needs to be carefully reviewed to avoid hallucinations, security issues, or accruing technical debt more generally. In our work, we have studied the question of whether developers actually treat LLM-generated code differently during code review. We conducted an eye tracking study with 32 software practitioners, measuring fixations and saccades when reviewing
LLM-labelled vs. unlabelled identical code.
Crucially, all code in our study was written by human developers: we employed a WoZ design in which no code was actually LLM-generated. This allows us to attribute any behavioural differences solely to developers' \textit{perception} of the code's origin, independent of the actual quality or style of any particular LLM. We find that reviewers indeed fixate on LLM-generated code for a longer time. However, this effect only becomes visible for longer or more complex code. We do not observe a meaningful difference in saccades, which is an operationalisation of how thoroughly the code is studied. Based on qualitative data and individual gaze paths, we observe important individual differences between participants with regard to how they perceive LLM-generated code and interact with the prompts.

We conclude that, while developers are often distrustful of LLM-generated code and spend more time reviewing it, this additional time does not translate into a more thorough review as measured via saccade lengths. This suggests that there is a need for specific policies and instructions for \emph{how} to review LLM-generated code, going beyond general guidance to be ``extra careful''. We also find that prompts (input to LLM code generation) can serve as a valuable aid during code review, helping reviewers verify alignment between intent and generated code. We recommend that prompts be treated as code metadata and made available on-demand within the development environment, rather than embedded as comments. Designing tools that support this form of prompt-to-code traceability without disrupting the review flow is an important direction for future work.

\section*{Data Availability Statement}
All material, scripts, and protocols are available in our replication package~\cite{replicationpackage}.
We cannot share the interview transcripts and raw data since they might break the anonymity of our participants or their organizations.

\section*{Acknowledgment}
We thank all participants for their time and willingness to take part in this study.
This work was partially supported by the Wallenberg AI, Autonomous Systems and Software Program (WASP) funded by the Knut and Alice Wallenberg Foundation, and by the HIVEMIND project funded by the European Union's Horizon Europe research and innovation (grant agreement 101189745). Icons used in illustrations are downloaded from flaticon.com.


\balance
\bibliographystyle{IEEEtran}
\bibliography{bib.bib}

\end{document}